# Evidence of multiband superconductivity in the *β*-phase Mo$_{1-x}$Re$_x$ alloys

Shyam Sundar[1], L S Sharath Chandra[2], M K Chattopadhyay[1,2] and S B Roy[1,2]

[1] Homi Bhabha National Institute at RRCAT, Indore, Madhya Pradesh 452 013, India
[2] Magnetic and Superconducting Materials Section, Raja Ramanna Center for Advanced Technology, Indore, Madhya Pradesh 452013, India

## Abstract

We present a detailed study of the superconducting properties in the *β*-phase Mo$_{1-x}$Re$_x$ ($x$ = 0.25 and 0.4) solid solution alloys pursued through magnetization and heat capacity measurements. The temperature dependence of the upper critical field $H_{C2}$(T) in these binary alloys shows a deviation from the prediction of the Werthamer–Helfand–Hohenberg (WHH) theory. The temperature dependence of superfluid density estimated from the variation of lower critical field $H_{C1}$ with temperature, cannot be explained within the framework of a single superconducting energy gap. The heat capacity also shows an anomalous feature in its temperature dependence. All these results can be reasonably explained by considering the existence of two superconducting energy gaps in these Mo$_{1-x}$Re$_x$ alloys. Initial results of electronic structure calculations and resonant photoelectron spectroscopy measurements support this possibility and suggest that the Re-5d like states at the Fermi level may not intermix with the Mo-5p and 5s like states in the *β*-phase Mo$_{1-x}$Re$_x$ alloys and contribute quite distinctly to the superconductivity of these alloys.

## Introduction

There has been a widespread interest recently in the multiband effect in superconductors, especially after the discovery of superconductivity in iron pnictide compounds [1]. Several new materials have been identified where the multiband effect is considered to be important [2–10]. Even before the recent interest, multiband effects were also known to govern the superconductivity in some well known materials like MgB2, NbSe2, borocarbides, Nb3Sn and MgCNi3 [11–15]. All these superconductors have complex crystallographic structures and complex Fermi surfaces. In this context it is interesting to note here that a two-band model was once considered for understanding the superconductivity in elemental body centered cubic (bcc) Niobium [16]. However, to the best of our knowledge the same framework has not been used so far to investigate the superconducting properties in other cubic metals and solid solution alloys. This is due to the fact that impurities, disorder and inter-band scattering can suppress the multiband effect in potential superconductors [2]. To this end, we re-investigate the interesting but not so well explored superconducting properties of *β*-phase Mo$_{1-x}$Re$_x$ binary solid solution alloys and show that various interesting superconducting properties of these alloys with bcc crystal structure can be explained within the framework of two superconducting gaps. Mo1-*x*Re*x* alloys possess excellent mechanical properties at elevated temperatures and find widespread applications in aerospace and defense industries, medical fields and welding production [17–20]. Superconductivity has been observed in Mo1-*x*Re*x* alloys across a wide solid-solution range of its phase diagram. The superconducting transition temperature $T_C$ in some of the alloy compositions is about an order of magnitude higher than the $T_C$ = 0.9 K in Mo and $T_C$ = 1 K in Re [21]. The $T_C$ of Mo1-$x$Re$x$ alloys varies non linearly with $x$ [22]. The Mo0.60Re0.40 alloy was identified as a strong coupling superconductor with a normalized energy gap $2/k_BT_C$ = 5.0. This was well above the value of 3.52 predicted by the Bardeen–Cooper–Schrieffer (BCS) theory of a weakly coupled superconductor [23]. Shum et al provided an explanation for the enhancement of $T_C$ in Mo0.60Re0.40 alloy by considering lattice softening [24]. The mass defect of Mo and Re i.e. $M$Re/$M$Mo = 1.94, disturbs the phonon spectrum and leads to the quasi local vibration or Brout–Visscher mode [25]. This mode contributes appreciably to the electron-phonon coupling function $\alpha^2F(\omega)$ and to $2/k_BT_C$. However, subsequent point contact spectroscopy studies by Tulina and Zaitsev pointed out that the enhancement of the electron-phonon coupling ($\lambda_{ep}$) from the lattice softening alone could not explain the enhancement of $T_C$ in Mo1-$x$Re$x$ alloys [26]. They argued that there must be a significant contribution from the electronic factor $N(0) <I^2>$ ($N(0)$ is the electron density of states and $<I^2>$ is the matrix element of the electron phonon coupling) towards the enhancement of $T_C$ in these alloys [26].

Molybdenum has an unoccupied $d$ band just above the Fermi level and it is quite clear that the addition of Re fills those unoccupied states and enhances the density of states (DOS) of the alloy system [21]. According to Matthias' empirical rule [21], the $T_C$ for solid solutions of transition metals shows its maxima at valence electrons per atom ($e/a$ ratio) around 4.7 and 6.4. The maximum in the $T_C$ also corresponds to the maximum in the Sommerfeld coefficient of electronic heat capacity $\gamma$. This indicates that the maximum in the $T_C$ is observed for the maximum electron density of states at the Fermi level [21]. However, the electron density of states for the solid solutions corresponding to $e/a$ = 6.4 is quite low as compared to that corresponding to $e/a$ = 4.7 and it is to be noted that Mo1-$x$Re$x$ solid solutions belong to the former regime [27]. Hence, the exact reason for the enhancement of $T_C$ in Mo1-$x$Re$x$ solid solutions still remains a matter of debate [24, 26].

The electronic properties of the Mo1-$x$Re$x$ alloys are also quite interesting and the existence of Fermi pockets and associated electronic topological transition (ETT) have been established in the Mo1-$x$Re$x$ alloys above the critical concentration $x_C$ = 0.11 through various experimental and theoretical studies [28–32]. The direct evidence of this ETT has been obtained recently with the help of angle resolved photoemission spectroscopy measurements along the H-N direction of the Brillouin zone [33]. However, any correlation between the ETT and the superconducting properties of the Mo1-$x$Re$x$ alloys is yet to be established. Apart from all these interesting microscopic properties, the Mo1-$x$Re$x$ alloys may also be useful for superconducting radio-frequency cavity applications [34].

In this paper we present a detailed study of temperature($T$) and magnetic field ($H$) dependence of magnetization ($M$) and heat capacity ($C$) in the Mo1-$x$Re$x$ ($x$ = 0.25 and 0.4) alloys. We show that the temperature dependence of $C(T)$ in these alloys can be explained by considering the existence of two superconducting gaps. A positive curvature is observed in the temperature dependence of the upper critical field $H_C2$, which is a possible signature of a multiband effect. The superfluid density (estimated from the temperature dependence of the lower critical field $H_C1$) can also be understood by considering the existence of two superconducting gaps. We note that this multiband effect in the Mo1-$x$Re$x$ alloys is observed in that alloy composition range where the appearance of the Fermi pockets above a critical value $x > x_C$ has earlier been reported in the literature [33].

## Experimental details

Polycrystalline samples of $Mo_{1-x}Re_x$, where ($x$ = 0.25, 0.40) were prepared by melting constituent elements with purity better than 99.95% in an arc furnace under 99.999% argon atmosphere. The samples were flipped and remelted six times to improve the homogeneity. Figure 1 shows the x-ray diffraction patterns of these alloys obtained with a Geigerflex diffractometer (Rigaku, Japan) which indicate that these samples have formed in the bcc phase (space group: Im3m). The lattice parameters obtained are about 3.135 ± 0.001 A0 and 3.126 ± 0.001 A0, respectively, for $x$ = 0.25 and $x$ = 0.40. The heat capacity measurements were performed in the temperature range 2–15 K in various applied magnetic fields up to 3 T using a Physical Property Measurement System (PPMS; Quantum Design, USA). The magnetization measurements were performed using a Superconducting Quantum Interference Device (SQUID) based Vibrating Sample Magnetometer (SQUID-VSM; Quantum Design, USA).

## Results and discussion

Figures 2(*a*) and (*b*) show the temperature dependence of heat capacity $C(T)$ for the Mo0.75Re0.25 and Mo0.60Re0.40 alloys, respectively, at various applied magnetic fields. The superconducting transition temperature $T_C$ is estimated as that temperature where the temperature derivative of the heat capacity is minimum. The estimated value of $T_C$ is 9.6±0.1 K for the Mo0.75Re0.25 alloy and 12.4±0.2 K for the Mo0.60Re0.40 alloy. The application of a magnetic field shifts the $T_C$ to lower temperatures. A field of 2 T suppresses the superconductivity to below 2 K in the Mo0.75Re0.25 alloy, whereas about 3 T is needed to achieve the same suppression in the Mo0.60Re0.40 alloy. In such a case the $C(T)$ in the normal state can be expressed by the functional form $C(T) = C_e + C_L$ where $C_e = \gamma T$ is the electronic contribution to heat capacity and $C_L = \beta T^3 + \delta T^5$ is the lattice contribution to heat capacity[35]. Figures 2(*c*) and (*d*) show the plots of $C/T$ versus $T^2$ of these alloys in the normal state obtained by applying high magnetic fields, which suppressed the superconducting transition temperature below 2 K. The $C/T$ is linear in $T^2$ just above 2 K. However, a deviation of $C/T$ from linearity appears at temperatures well below $T_C(H = 0)$ (blue dashed line in figures 2(*c*) and (*d*)). The temperature dependence of heat capacity $C(T)$ can be fitted with the functional form $\gamma T + \beta T^3 + \delta T^5$ (red solid line in figures 2(*c*) and (*d*)) over an extended temperature range well above $T_C(H = 0)$. The Debye temperature $\theta_D$ can be estimated from the coefficient $\beta$ as $\theta^3_D = 1943.66/\beta$. The estimated Debye temperature $\theta_D$ is 440 ± 4 K for the Mo0.75Re0.25 alloy and $\theta_D$ is 373 ± 2 K for the Mo0.60Re0.40 alloy. Morin and Maita [27] reported a $\theta_D$ value of 340 K for Mo0.60Re0.40 alloy, while $\theta_D$ value reported by Stewart and Giorgi [23] for the same alloy was 325 K. The value of Sommerfeld coefficient of electronic heat capacity $\gamma$ is estimated to be about 3.83 ± 0.02 mJ mol-1 K2 and 4.48 ± 0.02 mJ mol-1 K$^2$ fortheMo0.75Re0.25 andMo0.60Re0.40 alloys, respectively. The $\gamma$ value reported earlier for Mo0.60Re0.40 alloy agrees well with the present results [23].

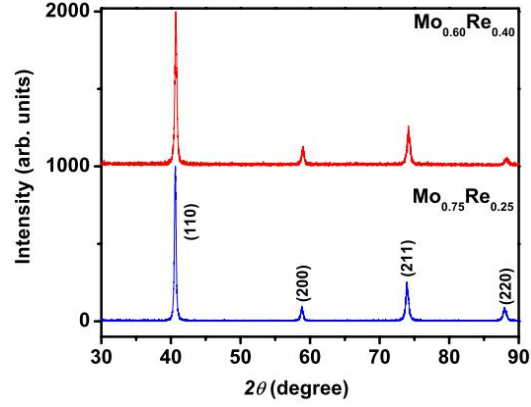

**Figure 1:** X-ray diffraction pattern for Mo0.75Re0.25 alloy and Mo0.60Re0.40 in the range 30–900 obtained using Cu-K$\alpha$ radiation. The most intense peak of each of the patterns is scaled to 1000 in order to present the patterns in the same scale. The intensity of x-ray diffraction pattern of Mo0.60Re0.40 alloy is shifted upwards by 1000 (for the clarity). The samples are found to have a bcc structure and space group: Im3m.

Figures 3(*a*) and (*b*) show the magnetization (*M*) as a function of magnetic field (*H*) at various temperatures below the $T_C$ of the Mo0.75Re0.25 and Mo0.60Re0.40 alloys, respectively. The insets show the expanded view of the field dependence of magnetization near $H_{C2}$. The upper critical field $H_C2$ is estimated from the magnetic field dependence of magnetization as the field at which the irreversible magnetization (giving rise to a hysteresis loop) reduces to zero. This field is slightly above the cross-over field from the diamagnetism to paramagnetism. These results are complemented with heat capacity measurements by noting down the temperature $T_C(H)$ of the jump in the $C(T)$ at various applied magnetic fields. Figures 3(*c*) and (*d*) show the magnetic field dependence of magnetization *M(H)* below $T_C$ of the Mo0.75Re0.25 and Mo0.60Re0.40 alloys, respectively, with an enlarged low *H* region. The measurements were performed after cooling the sample in the zero magnetic field to the desired temperature $T < T_C$ from well above $T_C$. The data have been corrected for the demagnetization effects. At low fields, the magnetization *M(H)* is linear in -*H* indicating that the sample is in the Meissner state. A procedure of a linear fit of a number of data points near $H_{appl} = 0$ after equating *M* to *H*, is used to estimate the demagnetization factor in these alloys. Then the effective magnetic field is estimated as $H_{eff} = H_{appl} - \alpha M$. The lower critical field $H_{C1}$, below which a type-II superconductor remains in the Meissner state, is in principle estimated from the deviation from the linearity in the low field *M* versus *H* plot. However, such estimation of $H_{C1}$ may be impaired by the Bean–Livingston surface barrier and/or geometrical barrier effects [36, 37]. In order to estimate the $H_{C1}$, a straight line is fitted to the *M-H* curve and the difference *M* between the measured magnetization and the fitted curve is estimated for a wide magnetic field region [37–39]. The $(M)^{1/2}$ is then plotted as a function of *H* and the value of $H_{C1}$ is estimated as the field at which a fitted straight line to this curve crosses the *H* axis [37–39]. We have observed that while this procedure is applicable for determining $H_{C1}$ in the Mo0.60Re0.40 alloy, $(M)^{1/2}$ is not linear in *H* for the Mo0.75Re0.25 alloy. Hence, for this latter alloy $HC1$ is estimated as the field at which the *M-H* curve deviates from linearity. Since $H_{C1}$ will be different for different criteria, we have estimated $HC1$ as that field at which the rise of $d^2M/dH^2$ at high fields extrapolates to zero. We have also crosschecked some of the estimated $HC1$ values in both the alloys following another procedure, which involves estimation of d*M*/d*H* from the measured isothermal *M(H)* curves both in increasing and decreasing cycle [40]. The temperature dependence of $H_{C2}$ is shown for the Mo0.75Re0.25 and Mo0.60Re0.40 alloys in figure 4. The derivative $(dH_{C2}/dT)_{T=T_C}$ estimated by fitting a straight line to the data points just below $T_C$ turns out to be about $-0.159 \pm 0.005$ T K-1 for the Mo0.75Re0.25 alloy and $-0.29 \pm 0.01$ T K-1 for the Mo0.60Re0.40 alloy. Within the framework of Werthamer, Helfand and Hohenberg (WHH) model [41], the temperature dependence of $H_{C2}$ can be expressed in the dirty limit as

$$\ln \frac{1}{t} = \sum_{\nu=-\infty}^{\infty} \left\{ \frac{1}{|2\nu+1|} - \left[ |2\nu+1| + \frac{\bar{h}}{t} + \frac{(\alpha_M \bar{h}/t)^2}{|2\nu+1| + (\bar{h}+\lambda_{SO})/t} \right]^{-1} \right\}, \quad [1]$$

where $t = T/T_C$, $\bar{h} = 2eH(v_f^2\tau/6\pi T_C) = (4/\pi^2) H_{C2}T_C/(-dH_{C2}/dT)_{T=T_C}$ with Fermi velocity $v_f$ and the relaxation time of electrons $\tau$, $\alpha_M = 3/2mv_f^2\tau = H_{C2}(0)/1.84\sqrt{2}\, T_C$ and $\lambda_{SO} = 1/3\pi T_C\tau_2$ with the relaxation time of electrons for spin-orbit interaction $\tau_2$. The temperature dependence of $H_{C2}$ estimated using the WHH model (dashed lines in figure 4) by taking experimentally obtained $(dH_{C2}/dT)_{T=T_C}$ matches with the experimental observations only at temperatures close to the $T_C$. This deviation from the WHH model indicates that the $H_{C2}(T)$ line in these alloys has a positive curvature. We have also tried to fit the $H_{C2}(T)$ over a large temperature range by taking $(dH_{C2}/dT)_{T=T_C}$ and $T_C$ as fitting parameters and the corresponding fit is shown as solid lines in figure 4. The fitted curve matches with the experimental data at low temperatures and deviates at temperatures close to $T_C$ for both the alloys. The values of $(dH_{C2}/dT)_{T=T_C}$ obtained are $-0.193 \pm 0.002$ T K-1 and $-0.335 \pm 0.005$ T K-1 for the Mo0.75Re0.25 and Mo0.60Re0.40 alloys, respectively. These values are comparatively higher than those estimated experimentally, which leads to a deviation at temperature close to $T_C$. The value of $T_C$ obtained as fitting parameter is $9.3 \pm 0.06$ K for the Mo0.75Re0.25 alloy and $12.2 \pm 0.1$ K for the Mo0.60Re0.40 alloy. These values are smaller than those observed experimentally. For both the alloys, the fitting parameter $\alpha_M$ lies between 0.08 to 0.18 and $\lambda_{SO}$ is about zero. This indicates that the paramagnetic effects and spin orbit interaction are negligible in these alloys. The values of the temperature dependent $H_{C2}$ for the present alloys are comparable to those reported earlier in the literature [42, 43]. Even in those earlier reports the temperature dependence of $H_{C2}$ for various $Mo_{1-x}Re_x$ solid solutions showed deviation from the predictions of theoretical model available at that time, namely the Abrikosov–Gorkov model [42].

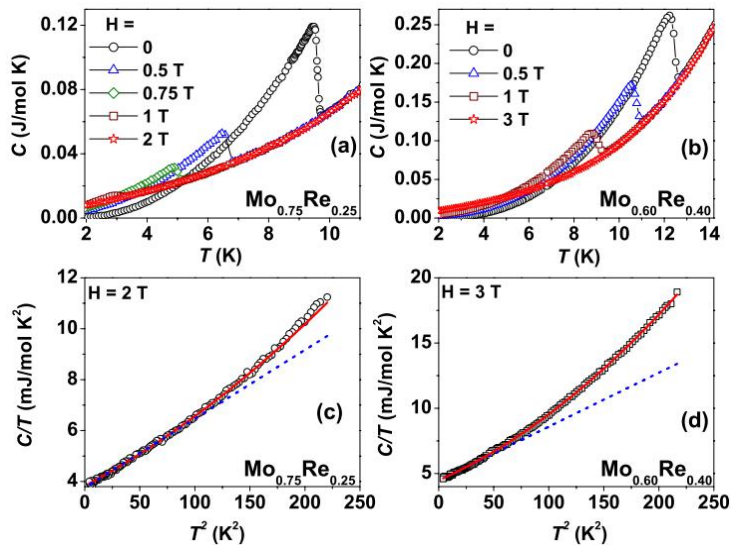

**Figure 2.** Temperature dependence of heat capacity of (*a*) Mo0.75Re0.25 and (*b*) Mo0.60Re0.40 alloys in different magnetic fields. The superconducting transition temperature $T_C$ is 9.6 ± 0.1 K and 12.4 ± 0.2 K, respectively, for the Mo0.75Re0.25 and Mo0.60Re0.40 alloys. The panels (*c*) and (*d*) present the $C/T$ versus $T^2$ of these alloys in the normal state. The open symbols are experimental points. The blue dashed line is the linear fit. The red solid line is the fit using $C(T) = \gamma T + \beta T^3 + \delta T^5$.

The temperature dependence of $H_{C1}$ (see figure 5) for the Mo0.75Re0.25 and Mo0.60Re0.40 alloys shows the usual form $H_{C1}(T) = H_{C1}(0)[1 - (T/T_C)^2]$ down to the lowest temperature [44]. The fit using the above equation at low temperatures yields $H_{C1}(0) = 68.5 \pm 0.1$ mT for Mo0.75Re0.25 and $H_{C1}(0) = 81.4 \pm 0.1$ mT for Mo0.60Re0.40. The values of $H_{C1}(T)$ for the Mo0.60Re0.40 alloy are comparable to those of the Mo0.64Re0.36 alloy reported earlier in the literature [43]. However, the temperature dependence of $H_{C1}$ for the Mo0.75Re0.25 alloy is different from that reported for the same composition [42].

For a superconductor in the local limit with $\xi(0) \ll \lambda$ (where $\xi(0)$ and $\lambda$ are coherence length and penetration depth, respectively), the normalized super fluid density $\rho_s(T)$ in the framework of local London model is given by [45, 46]

$$\rho_s(T) = \frac{\lambda^2(0)}{\lambda^2(T)} = \frac{H_{C1}(T)}{H_{C1}(0)}. \qquad [2]$$

The Figure 6(a) and (b) show the temperature dependence of $H_{C1}(T)/H_{C1}(0)$ of the Mo0.75Re0.25 and Mo0.60Re0.40 alloys, respectively, which represent the superfluid density in these alloys. The open symbols are the experimental data points. For a single gap superconductor, the normalized superfluid density can be expressed as [47]

$$\rho_s(T) = 1 + 2\int_{\Delta(T)}^{\infty} \frac{dF(E)}{dE} D(E) dE \qquad [3]$$

where $F(E)$ is the Fermi function and $D(E) = \frac{E}{\sqrt{E^2 - \Delta(T)^2}}$. Here, $\Delta(T)$ is the superconducting gap [45, 47]. For an isotropic superconductor, $\Delta(T)$ is given by $\Delta(T) = \Delta(0) \tanh\{1.82[1.018(T_C/T - 1)]^{0.51}\}$ where $\Delta(0)$ is the superconducting gap at absolute zero [48].

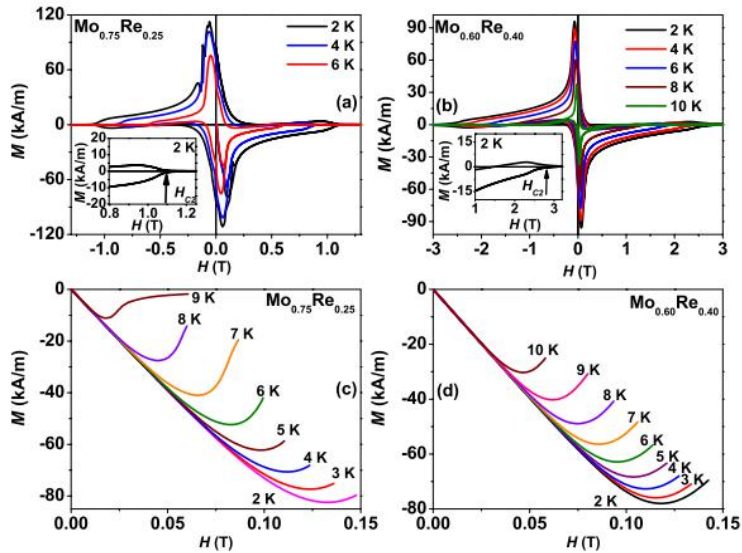

**Figure 3.** Magnetic field dependence of magnetization of (*a*) Mo0.75Re0.25 and (*b*) Mo0.60Re0.40 below *T*C. The insets show the expanded view of the field dependence of magnetization near *H*C2. The upper critical field *H*C2 is estimated from the magnetic field dependence of magnetization as the field at which the irreversible magnetization (giving rise to a hysteresis loop) in the isothermal *M-H* curves reduces to zero. The panels (*c*) and (*d*) present the magnetic field dependence of magnetization at various temperatures below $T_C$ of the Mo0.75Re0.25 and the Mo0.60Re0.40 alloys, respectively, in low *H* regime. Magnetization results presented here are in the form of closely spaced data points.

The dotted lines in figures 6(*a*) and (*b*) show the temperature dependence of normalized superfluid density estimated using the equation (3) for an isotropic single gap superconductor with $\Delta(0)$ = 5.5±0.5 K for the Mo0.75Re0.25 alloy and $\Delta(0)$ = 20.5 ± 0.4 K for the Mo0.60Re0.40 alloy, respectively. The goodness of fit is estimated from the Pearson's $\chi^2$ test method as $\chi^2 = \Sigma_{i=1}^{n} \frac{(O_i - E_i)^2}{E_i}$ where $O_i$ is the experimental value, $E_i$ is expected or the theoretical value and *n* is the number of data points. The value of $\chi^2$ is about 0.19 for the Mo0.75Re0.25 alloy and 0.26 for the Mo0.60Re0.40 alloy. The estimated theoretical curve matches well with the experimental data at high temperatures. However, marked deviation observed at low temperatures indicates the possibility of the existence of two superconducting gaps [49] or the presence of a single anisotropic gap [50]. For a two gap superconductor, the normalized superfluid density can be expressed as [47].

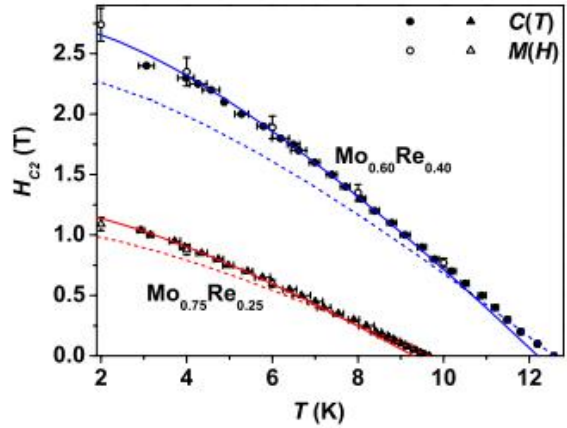

**Figure 4.** Temperature dependence of the upper critical field $H_{C2}$ for the Mo1-xRex alloys. The dashed lines are the fit using Werthamer, Helfand and Hohenberg (WHH) model by taking experimentally obtained (d*H*C2/d*T* )*T* =*T*C. The fit matches with the experimental observations only at temperatures close to *T*C The solid lines represent the fits to the data by taking (d*H*C2/d*T* )*T* =*T*C and *T*C as fitting parameters. In this case, the experimental data points deviate from WHH model at temperatures close to *T*C.

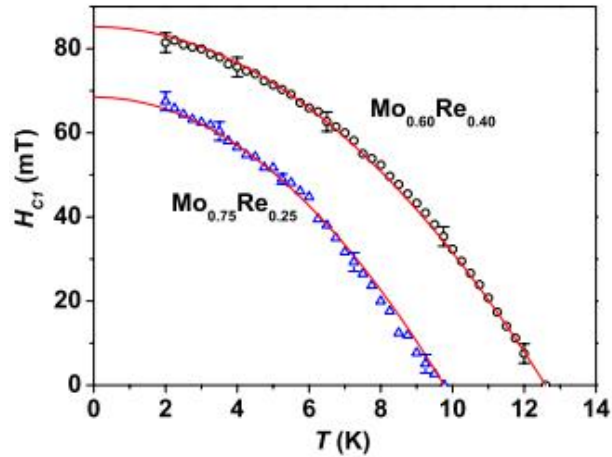

**Figure 5.** Temperature dependence of the lower critical field *H*C1 for Mo1-xRex binary alloys. The solid lines represent the fits to the data using the form *H*C1(*T* ) = *H*C1(0) [1 - (*T* /*T*C)²].

$$\rho_s(T) = 1 + 2 \left( c \int_{\Delta_S(T)}^{\infty} \frac{dF(E)}{dE} D_S(E) dE + (1-c) \int_{\Delta_L(T)}^{\infty} \frac{dF(E)}{dE} D_L(E) dE \right) \quad [4]$$

where $\Delta_S$ and $\Delta_L$ are the small and large superconducting gap, respectively. The parameter $c$ is the fraction that the small gap contributes to the superconductivity. At low temperatures ($T/T_C < 0.5$) where $\gamma(T)$ varies within 10% of $\Delta(0)$, the equation (4) reduces to

$$\rho_s(T) = 1 - c \left( \frac{2\pi \Delta_S(0)}{k_B T} \right)^{1/2} \exp\left( -\frac{\Delta_S(0)}{k_B T} \right) \\ - (1-c) \left( \frac{2\pi \Delta_L(0)}{k_B T} \right)^{1/2} \exp\left( -\frac{\Delta_L(0)}{k_B T} \right) \quad [5]$$

The fit to the temperature dependence of superfluid density at low temperatures using equation (5) can distinguish between the presence of a single anisotropic gap and two superconducting gaps. In case of the presence of a single anisotropic gap, the parameter $c$ in equation (5) will approach unity or zero. Any one of the $\Delta_S(0)$ and $\Delta_L(0)$ should also approach zero and the other should have a value less than $1.76 k_B T_C$ [51]. If the system has two superconducting gaps, then the parameter $c$ will have a value such that $0 < c < 1$ and both the $\Delta_S(0)$ and $\Delta_L(0)$ will have non zero values. The insets to the figures 6(*a*) and (*b*) show the fit to the superfluid density at low temperatures using equation (5) for the Mo0.75Re0.25 and Mo0.60Re0.40 alloys, respectively. The fits indicate the existence of two superconducting gaps in these alloys and negate the possibility of a single anisotropic gap. The value of $\Delta_L(0)$ is very close to the BCS theoretical limit of $1.76 k_B T_C$. Hence, we have used equation (4) to fit the superfluid density in whole temperature range (red solid lines in the figures 6(*a*) and (*b*)) by considering two isotropic superconducting gaps. The $\chi^2$ is about 0.13 for the Mo0.75Re0.25 alloy and 0.0074 for the Mo0.60Re0.40 alloy. Note that $\chi^2$ values are smaller for two gap models as compared to that for single gap models. The values of $\gamma\Delta_L(0)$ ($\Delta S(0)$) = 18.0 ± 0.6 K (9.0 ± 0.6 K) for the Mo0.75Re0.25 and $\Delta_L(0)$ ($\Delta S(0)$ = 22.5± 0.6 K (6.0± 0.5 K) for the Mo0.60Re0.40 are slightly higher (quite lower) than the BCS limit of $1.76 k_B T_C$. The estimated value of $c$ is about 25 ± 1% and 12 ± 1% for the Mo0.75Re0.25 and Mo0.60Re0.40 alloys, respectively.

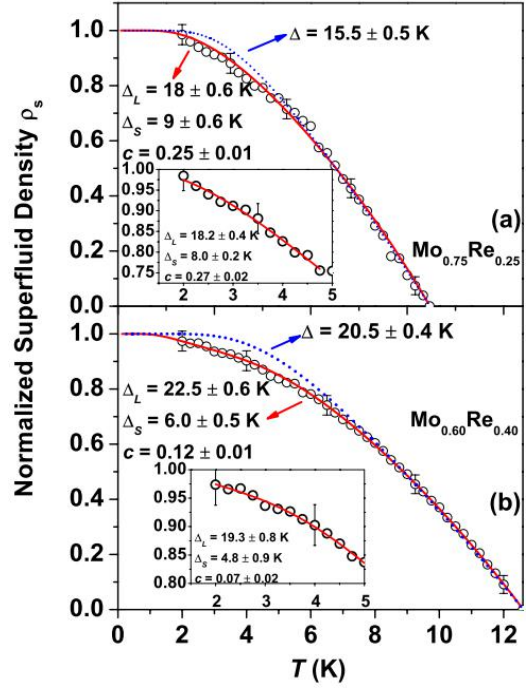

**Figure 6.** Temperature dependence of superfluid density of (*a*) Mo0.75Re0.25 and (*b*) Mo0.60Re0.40. The open symbols are experimental data points, the dotted line is the fit obtained by considering single gap and the solid line is the fit obtained by considering two superconducting gaps. The insets show the fit below $T < 0.5T_C$ using equation (5) to know whether the system has two superconducting gaps or an anisotropic gap. The analysis shows that the temperature dependence of superfluid density of these alloys can be explained only after considering the existence of two superconducting gaps.

Additional evidence for the existence of two superconducting gaps in the present Mo1-*x*Re*x* alloys can be obtained directly from the temperature dependence of heat capacity in the superconducting state. The electronic heat capacity in the superconducting state $C_S$ in the zero magnetic field is estimated by subtracting the contribution of the lattice heat capacity $C_L$ from the total heat capacity and is plotted as $C_S/\gamma T_C$ in figures 7(*a*) and (*b*). The values of heat capacity jump at $T_C$, $\Delta C_S/\gamma T_C$ are about 1.7 and 2 for the Mo0.75Re0.25 and Mo0.60Re0.40 alloys, respectively and these are substantially higher than the BCS value of 1.43 for the weak coupling superconductors. This again suggests that the superconductivity in the present binary Mo1-*x*Re*x* alloys is rather unconventional.

The electronic heat capacity in the superconducting state for a superconductor with two superconducting gaps corresponding to two bands without interband scattering is given by [52]

$$C_S/\gamma T_C = cC_{S1}/\gamma_1 T_C + (1-c)C_{S2}/\gamma_2 T_C, \qquad [6]$$

where $C_{Si}$ (*i* = 1, 2) corresponds to heat capacity resulting from superconducting gap $\Delta_i$ and $c = \gamma_1/\gamma$ is the fraction that the small gap contributes to the superconductivity and $\gamma = \gamma_1 + \gamma_2$. Here $\gamma_1$ ($\gamma_2$) is the normal state $\gamma$ for the band 1 (band 2) that is superconducting.
Here, the $C_{Si}/\gamma_i T_C$ is given by [53, 54]

$$C_{Si}(T)/\gamma_i T_C = \frac{6\alpha_i^2}{\pi^2} \frac{1}{4\pi} \frac{T_C}{T} \int_0^{2\pi} d\phi \int_0^{\pi} d\theta \sin\theta \qquad [7]$$

$$\int_0^{\infty} dx \left(-\frac{df_i}{dE_i}\right) \left(E_i^2 - \frac{T}{2}\frac{d\delta_i^2}{dT}\right)$$

where $E_i = x^2 + \delta_i^2$, $f_i = (1 + \exp(\alpha_i T_C E_i/T))^{-1}$ and $\alpha_i = \Delta_i(0)/k_B T_C$. The $\Delta_i(0)$ is the superconducting gap at absolute zero. For an isotropics wave superconductor, $\delta_i = \Delta_i(T)/\Delta_i(0)$, where $\Delta_i(0)$ is a constant, $\delta_i = (\Delta_i(T)/\Delta_i(0)) \cos n\varphi$ for line nodes and $\delta_i = (\Delta_i(T)/\Delta_i(0)) \sin n\theta$ for point nodes, where $\theta$ and $\varphi$ are the polar and azimuthal angles over the Fermi surface. The equation (6) reduces to a single gap model when $c = 0$.

The dotted blue lines in figures 7(*a*) and (*b*) represent the temperature dependence of heat capacity in the superconducting state with a single isotropic superconducting gap $\Delta(0) = 19.0 \pm 0.5$ K for the Mo0.75Re0.25 alloy and $\Delta(0) = 26.5 \pm 0.5$ K for the Mo0.60Re0.40 alloy, respectively. The goodness of fit $\chi^2$ is about 0.1 for the Mo0.75Re0.25 alloy and 0.114 for the Mo0.60Re0.40 alloy. However, at low temperatures, the value of $C_S/\gamma T_C$ obtained experimentally is higher than that corresponding to the model fitting using single isotropic gap. We have also observed that the model fitting by considering a single anisotropic gap (not shown here for the sake of clarity) cannot explain the temperature dependence of the heat capacity in these Mo1-*x*Re*x* alloys. Then we have fitted our experimental results using equations (6) and (7) (solid red line in figures 7(*a*) and (*b*)) and found that the two isotropic superconducting gaps can explain the temperature dependence of heat capacity in the superconducting state. The $\chi 2$ value is about 0.041 for the Mo0.75Re0.25 alloy and 0.013 for the Mo0.60Re0.40 alloy. Similar to the fitting of temperature dependence of superfluid density, the $\chi 2$ values corresponding to the fitting of the temperature dependence of heat capacity are also smaller for two gap models as compared to that for single gap model. The analysis shows that the value of the larger (smaller) of the two gaps $\Delta_L(0)$ ($\Delta_S(0)$) = 20.0±0.6 K (9.7 ± 0.5 K) for the Mo0.75Re0.25 alloy and $\Delta_L(0)$ ($\Delta_S(0)$) = 26.5 ± 0.6 K (8.2 ± 0.6 K) for the Mo0.60Re0.40 alloy is higher (lower) than the BCS limit of $1.76 k_B T_C$. The contribution from the smaller gap is about 10 ± 1% in the Mo0.75Re0.25 alloy whereas it is about 2.0 ± 0.2% in the Mo0.60Re0.40 alloy. These values, however, are relatively less as compared to that estimated from superfluid density. Such behavior has been observed earlier in another superconductor PrPt4Ge12 [46, 55]. This is probably due to the fact that the superfluid density estimated from $H_{C1}$ is a local property whereas heat capacity is a bulk property. We have also not considered the effect of inter-band scattering in analyzing the temperature dependence of heat capacity in the superconducting state. It is to be noted here that in an earlier study of the temperature dependence of electronic heat capacity in Mo0.60Re0.40 alloy, a clear deviation from the exponential behavior (corresponding to single energy gap) was indeed observed [23] but not analyzed further.

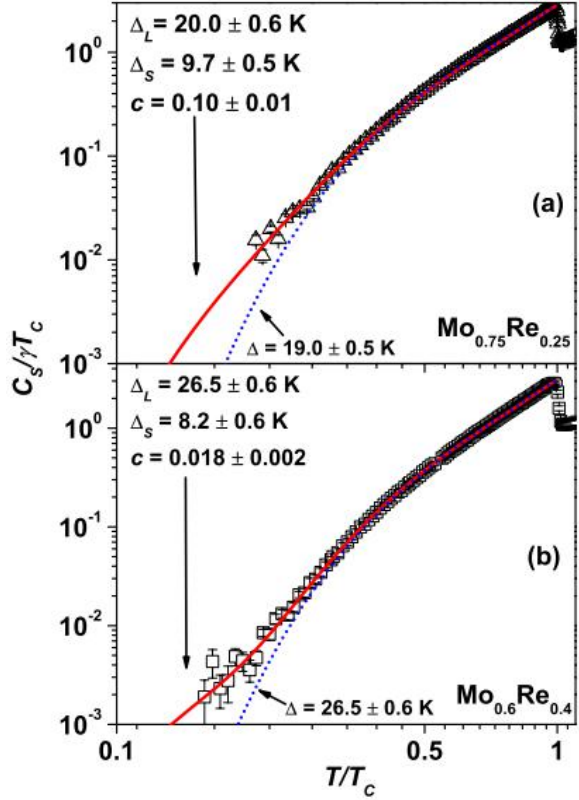

**Figure 7.** Temperature dependence of the electronic heat capacity in the superconducting state $C_S/\gamma T_C$ plotted as a function of $T/T_C$ for (*a*) Mo0.75Re0.25 and (*b*) Mo0.60Re0.40. The lines are fits to the experimental data (open symbols). The analysis shows that the temperature dependence of the heat capacity in these alloys can be explained by considering the existence of two superconducting gaps.

In the case of two-gap superconductors, the magnetic field dependence of the electronic part of heat capacity $C_S/T$ at temperatures well below $T_C$ should show two distinct linear regions with a change of slope at intermediate fields [2]. We have shown in figure 8, the plot of $C_S/T$ as a function of $H/H_{C2}$ at 2 K for both the Mo0.75Re0.25 and Mo0.60Re0.40 alloys. The $C_S/T$ corresponding to Mo0.60Re0.40 alloy is shifted upwards for clarity. In case of the Mo0.75Re0.25 alloy, the slope of the low field linear portion is slightly higher as compared to that at high fields, which is similar to other two gap superconductors such as MgB2 and NbSe2 [56]. Such behaviour is observed when the smaller of the two gaps vanishes at low fields and the corresponding normal electron contribution increases. However, in the case of the Mo0.60Re0.40 alloy, the change in the slope is quite subtle and is also reversed as compared to that of the Mo0.75Re0.25 alloy. This may be due to the enhanced inter-band scattering [2] in the Mo0.60Re0.40 alloy. The present *β*-phase Mo1-*x*Re*x* binary alloys have the bcc crystal structure, which is analogous to the elemental molybdenum. In this structure, the Mo and Re atoms randomly occupy the corner of the cube (0, 0, 0) and the body center (0.5, 0.5, 0.5). Thus, the presence of two superconducting gaps in these alloys at the first sight is quite surprising. However, the concentration of Re in the present alloys is higher than the critical concentration $x_C$ = 0.11 at which the existence of electronic topological transition in *β* -phase Mo1-*x*Re*x* binary alloys has been reported [28–33]. For the Mo1-*x*Re*x* alloys above $x_C$, a band crosses the Fermi level along the H-N direction of the Brillouin zone [33]. Initial results of our band structure calculations and resonant photoelectron spectroscopy experiments reveal that the there is a charge transfer from Re to Mo when Re is alloyed with the Mo [57]. Our study also reveals that there is a substantial change in the structure of density of states in Mo1-xRex alloys just below the Fermi level [57]. The density of states at the Fermi level of the Mo1-xRex alloys are mainly derived from the narrow Re $5d$ like states and the broad Mo$5p$ as well as Mo$5s$ like states. The Re $5d$ like states are not intermixed with the Mo $5p$ like and Mo $5s$ like states. These initial results [57] when compared with the results of angle resolved photoemission studies reported in literatures [33], indicate that the Re $5d$ like states can be linked to the band that crosses the Fermi level along the H-N direction of the Brillouin zone when Re is alloyed with Mo. It is natural to expect that the Fermi velocity in these narrow Re $5d$ like states is distinctly different from that in the broad Mo $5p$ like and Mo $5s$ like states. Therefore, we conjecture that these narrow Re $5d$ like states contribute to the superconductivity separately from the broad Mo $5p$ like and Mo $5s$ like states. It is also to be noted that the multiband superconductivity at the electronic topological transition has been observed in systems such as URhGe [58] and the high temperature superconducting pnictides [59].

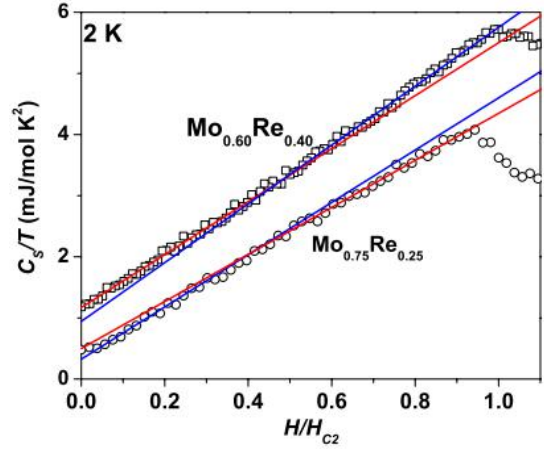

**Figure 8.** Magnetic field dependence of $C_S/T$ at 2 K as a function of $H/H_{C2}$ for the $Mo_{0.75}Re_{0.25}$ and $Mo_{0.60}Re_{0.40}$ alloys. The $C_S/T$ corresponding to $Mo_{0.60}Re_{0.40}$ alloy is shifted upwards for clarity. The magnetic field dependence of the heat capacity shows two linear regions and a change in slope at intermediate magnetic fields in these alloys.

## **Conclusion**

In summary, we have observed various anomalous features in the superconducting properties, namely the upper critical field and superfluid density of the $\beta$-phase $Mo_{1-x}Re_x$ ($x = 0.25$ and $0.4$) alloys. These anomalous features are suggestive of the existence of two superconducting gaps in these binary alloy superconductors. Further support for the presence of a multiband effect is obtained from the temperature dependence of the heat capacity in the superconducting state. At first sight, the possibility of such multiband effects in these $Mo_{1-x}Re_x$ alloys with relatively simple crystal structures is quite surprising. However, there are reports in the literature [28–33] which suggest the existence of an ETT in the Mo1-xRex alloys with the critical concentration $x_C = 0.11$. In this direction, preliminary results [57] of our electronic structure calculation and resonance photoelectron spectroscopy experiments in the present $Mo_{1-x}Re_x$ ($x = 0.25$ and $0.4$) alloys reveal the existence of narrow Re $5d$ like states and the broad Mo $5p$ as well as Mo $5s$ like states at the Fermi level, which contribute to a large enhancement in the density of states at the Fermi level. These narrow Re *5d* like states along with the broad Mo *5p* and Mo *5s* like states are possibly the source of the multiband effect in the present $Mo_{1-x}Re_x$ alloys.

## Acknowledgments

We would like to thank R K Meena for help in sample preparation, V S Tiwari and Gurvinderjit Singh for the x-ray diffraction measurements and V Ganesan and D Venkateshwaralu for some of the heat capacity measurements.

## References:

[1] Hunte F, Jaroszynski J, Gurevich A, Larbalestier D C, Jin R, Sefat A S, McGuire M A, Sales B C, Christen D K and Mandrus D 2008 *Nature* **453** 903
[2] Zehetmayer M 2013 *Supercond. Sci. Technol.* **26** 043001
[3] Kittaka S, Aoki Y, Shimura Y, Sakakibara T, Seiro S, Geibel C, Steglich F, Ikeda H and Machida K 2014 *Phys. Rev. Lett.* **112** 067002
[4] Zocco D A, Grube K, Eilers F, Wolf T and Lohneysen H V 2013 *Phys. Rev. Lett.* **111** 057007
[5] Wang H, Dong C, Mao Q, Khan R, Zhou X, Li C, Chen B, Yang J, Su Q and Fang M 2013 *Phys. Rev. Lett.* **111** 207001
[6] Seyfarth G, Brison J P, Knebel G, Aoki D, Lapertot G and Flouquet J 2008 *Phys. Rev. Lett.* **101** 046401
[7] Lortz R, Viennois R, Petrovic A, Wang Y, Toulemonde P, Meingast C, Koza M M, Mutka H, Bossak A and Miguel A S 2008 *Phys. Rev.* B **77** 224507
[8] Singh Y, Martin C, Bud'ko S L, Ellern A, Prozorov R and Johnston D C 2010 *Phys. Rev.* B **82** 144532
[9] Petrovic A P *et al* 2011 *Phys. Rev. Lett.* **106** 017003
[10] Kuroiwa S, Saura Y, Akimitsu J, Hiraishi M, Miyazaki M, Satoh K H, Takeshita S and Kadono R 2008 *Phys. Rev. Lett.* **100** 097002
[11] Pickett W 2002 *Nature* **418** 733
[12] Yokoya T, Kiss T, Chainani A, Shin S, Nohara M and Takagi H 2001 *Science* **294** 2518
[13] Shulga S V, Drechsler S-L, Fuchs G, Muller K-H, Winzer K, Heinecke M and Krug K 1998 *Phys. Rev. Lett.* **80** 1730
[14] Guritanu V, Goldacker W, Bouquet F, Wang Y, Lortz R, Goll G and Junod A 2004 *Phys. Rev.* B **70** 184526
[15] Walte A, Fuchs G, Muller K-H, Handstein A, Nenkov K, Narozhnyi V N, Drechsler S-L, Shulga S, Schultz L and Rosner H 2004 *Phys. Rev.* B **70** 174503
[16] Carlson J R and Satterthwaite C B 1970 *Phys. Rev. Lett.* **24** 461
[17] Wardsworth J and Wittenauer J P 1993 *Evolution of Refractory Metals and Alloys* ed E N C Dalder *et al* (Warrendale, OH: The Minerals, Metals and Materials Society)
[18] Heenstand R L 1993 *Evolution of Refractory Metals and Alloys* ed E N C Dalder *et al* (Warrendale, OH: The Minerals, Metals and Materials Society)
[19] Mannheim R L and Garin J L 2003 *J. Mater. Process. Technol.* **143–4** 533
[20] Mao P, Han K and Xin Y 2008 *J. Alloys Compounds* **464** 190
[21] Vonsovsky S V, Izyumov Yu A and Kurmaev E Z 1982 *Superconductivity of Transition Metals: Their Alloys and Compounds* (Berlin: Springer) (Engl. transl.)
[22] Ignat'eva T A and Cherevan' Yu A 1980 *Pis. Zh. Eksp. Teor. Fiz.* **31** 389
[23] Stewart G R and Giorgi A L 1978 *Solid State Commun.* **28** 969 [24] Shum D P,

Bevolo A, Staudenmann J L and Wolf E L 1986 *Phys. Rev. Lett.* **57** 2987
[25] Brout R and Visscher W 1962 *Phys. Rev. Lett.* **9** 54
[26] Tulina N A and Zaitsev S V 1993 *Solid State Commun.* **86** 55
[27] Morin F J and Maita J P 1963 *Phys. Rev.* **129** 1115
[28] Velikodny A N, Zavaritskii N V, Ignat'eva T A and Yurgens A A 1986 *Pis. Zh. Eksp. Teor. Fiz.* **43** 597
[29] Gornsoostyrev Y N, Katsnelson M I, Peschanskikh G V and Trefilov A V 2011 *Phys. Status Solidi* B **164** 185
[30] Skorodumova N V, Simak S I, Blanter Y M and Vekilov Y K 1994 *Pis. Zh. Eksp. Teor. Fiz.* **60** 549
[31] Ignat'eva T A and Velikodny A N 2002 *Low Temp. Phys.* **28** 403
[32] Ignat'eva T A 2007 *Phys. Solid State* **49** 403 Ignat'eva T A 2007 *Fiz. Tverd. Tela* **49** 389
[33] Okada M, Rotenberg E, Kevan S D, Schafer J, Ujfalussy B, Stocks G M, Genatempo B, Bruno E and Plummer E W 2013 *New J. Phys.* **15** 093010
[34] Andreone A, Barone A, Chiara A D, Fontana F, Mascolo G, Palmieri V, Peluso G, Pepe G and Scotti D U U 1989 *J. Supercond.* **2** 493
[35] Tari A 2003 *The Specific Heat of Matter at Low Temperatures* (London: Imperial College Press)
[36] Lyard L *et al* 2004 *Phys. Rev.* B **70** 180504
[37] Roy S B, Myneni G R and Sahni V C 2008 *Supercond. Sci. Technol.* **21** 065002
[38] Moshchalkov V V, Henry J Y, Marin C, Rossat-Mignod J and Jacquot J F 1991 *Physica* C **175** 407
[39] Liang R, Dosanjh P, Bonn D A, Hardy W N and Berlinsky A J 1994 *Phys. Rev.* B **50** 4212
[40] Liang R, Bonn D A, Hardy W N and Broun D 2005 *Phys. Rev. Lett.* **94** 117001
[41] Werthamer N R, Helfand E and Hohenberg P C 1966 *Phys. Rev.* **147** 295
[42] Joiner W C H and Blaugher R D 1964 *Rev. Mod. Phys.* **36** 67
[43] Lerner E, Daunt J G and Maxwell E 1967 *Phys. Rev.* **153** 487
[44] French R A 1968 *Cryogenics* **8** 301
[45] Ren C, Wang Z S, Luo H Q, Yang H, Shan L and Wen H H 2008 *Phys. Rev. Lett.* **101** 257006
[46] Sharath Chandra L S, Chattopadhyay M K and Roy S B 2012 *Phil. Mag.* **92** 3866
[47] Kim M S, Skinta J A, Lemberger T R, Kang W N, Kim H J, Choi E M and Lee S I 2002 *Phys. Rev.* B **66** 064511
[48] Carrington A and Manzano F 2003 *Physica* C **385** 205
[49] Maisuradze A, Schnelle W, Khasanov R, Gumeniuk R, Nicklas M, Rosner H, Leithe-Jasper A, Grin Y, Amato A and Thalmeier P 2010 *Phys. Rev.* B **82** 024524
[50] Maisuradze A, Nicklas M, Gumeniuk R, Baines C, Schnelle W, Rosner H, Leithe-Jasper A, Grin Yu and Khasanov R 2009 *Phys. Rev. Lett.* **103** 147002
[51] Okamoto H, Taniguti H and Ishihara Y 1996 *Phys. Rev.* B **53** 384
[52] Bouquet F, Wang Y, Fisher R A, Hinks D G, Jorgensen J D, Junod A and Phillips N E 2001 *Europhys. Lett.* **56** 856
[53] Padamsee H, Neighbor J E and Shiffman C A 1973 *J. Low Temp. Phys.* **12** 387
[54] Huang C L, Lin J-Y, Sun C P, Lee T K, Kim J D, Choi E M, Lee S I and Yang H D 2006 *Phys. Rev.* B **73** 012502
[55] Zhang J L *et al* 2013 *Phys. Rev.* B **87** 064502
[56] Boaknin E *et al* 2003 *Phys. Rev. Lett.* **90** 117003
[57] Shyam S *et al* 2014 unpublished
[58] Yelland E A, Barraclough J M, Wang W, Kamenev K V and Huxley A D 2011

*Nat. Phys.* **7** 890

[59] Innocenti D, Poccia N, Ricci A, Valletta A, Caprara S, Perali A and Bianconi A 2010 *Phys. Rev.* B **82** 184528